\documentclass [11pt]{article}
\usepackage[]{amsmath,amssymb,amsfonts,latexsym,amsthm,enumerate,paralist}
\usepackage[numeric,initials,nobysame]{amsrefs}

\long\def\symbolfootnote[#1]#2{\begingroup%
\def\thefootnote{\fnsymbol{footnote}}\footnote[#1]{#2}\endgroup}

\def\nred{n_{\textrm{r}}}
\def\nblue{n_{\textrm{b}}}

\date{}
\title{Optimal Bi-Valued Auctions}
\author{ {Oren Ben-Zwi\thanks{Faculty of Industrial Engineering and Management, Technion,
Haifa, Israel. Email: {\tt orenb@technion.ac.il}. This research was supported in part by the Google Inter-university center for Electronic Markets and Auctions.}} \and
{Ilan Newman\thanks{Department of Computer Science, Haifa University,
Haifa, Israel. Email: {\tt ilan@cs.haifa.ac.il}.}}}
\newtheorem{theorem}{Theorem}[section]
\newtheorem{lemma}[theorem]{Lemma}
\newtheorem{claim}[theorem]{Claim}
\newtheorem{definition}{Definition}

\renewcommand{\epsilon}{\varepsilon}

\newtheoremstyle{upright}%
        {8pt plus2pt minus4pt}%
        {8pt plus2pt minus4pt}%
        {\upshape}%
        {}%
        {\bfseries}%
        {:}%
        {1em}%
        {}%
\theoremstyle{upright}

\newtheorem*{informalremark}{Informal Remark}

\newtheorem*{example}{Example}

\newcommand{\ignore}[1]{}
\begin{document}

\maketitle

\begin{abstract}
We investigate \emph{bi-valued} auctions in the digital good setting and construct
an explicit polynomial time deterministic auction. We prove an unconditional tight
lower bound which holds even for random superpolynomial auctions.
The analysis of the construction uses the adoption of the finer
lens of \emph{general competitiveness} which considers additive losses on top
of multiplicative ones.  The result implies that general competitiveness
is the right notion to use in this setting, as this optimal auction is uncompetitive with respect
to competitive measures which do not consider additive losses.
\ignore{
We investigate \emph{bi-valued} auctions in the digital good setting and construct
an explicit optimal deterministic auction. The analysis of this construction uses the adoption of the finer
lens of \emph{general competitiveness} which considers additive losses on top
of multiplicative ones.  The result implies that general competitiveness
is the right notion to use in this setting, as this optimal auction is uncompetitive with respect
to competitive measures which do not consider additive losses.
}
\end{abstract}

\section{introduction}

Marketing a digital good may suffer from a low revenue due to incomplete
knowledge of the marketer. Consider, for example, a major sport event
with some $10^8$ potential TV viewers. Assume further that every potential
viewer is willing to pay $\$10$ or more to watch the event, and that no more than $10^6$
are willing to pay $\$100$ for that. If the concessionaire will charge $\$1$
or $\$100$ as a fixed pay per view price for the event, the overall collected
revenue will be $\$10^8$ at the most. This is worse than the revenue that can
be collected, having known the valuations beforehand.

This lack of knowledge motivates the study of \emph{unlimited supply, unit demand, single
item} auctions. Goldberg~et~al.~\cite{GoldbergHaWr2001} studied these auctions;
in order to obtain a prior free, worst case analysis framework, they suggested to compare
the revenue of these auctions to the revenue of optimal fixed price auctions.
They adopted the \emph{online algorithms} terminology~\cite{SleatorTa1985}
and named the revenue of the fixed price auction the \emph{offline revenue} and
the revenue of a multi price \emph{truthful} auction, i.e., an auction for
which every bidder has an incentive to bid its own value, \emph{online revenue}.
The \emph{competitive ratio} of an auction for a bid vector $b$ is defined to be
the ratio between the best offline revenue for $b$ to the revenue of that auction
on $b$. The competitive ratio of an auction is just the worst competitive ratio
of that auction on all possible bid vectors. For random auctions, a
similar notion is defined by taking the expected revenue. If an auction has a
constant competitive ratio it is said to be \emph{competitive}. If an auction has
a constant competitive ratio, possibly with some small additive loss, it is said to be
\emph{general competitive} (see Section~\ref{prel} for definitions). We remark that
later, Koutsoupias and Pierrakos~\cite{KoutsoupiasPi2010} used \emph{online auctions}
in the usual context of online algorithms, but here we will stick to
Goldberg~et~al.'s~\cite{GoldbergHaWr2001} notion.

This attempt to carefully select the right benchmark in order to obtain a prior free,
worst case analysis was a posteriori justified when Hartline and Roughgarden defined 
a general benchmark for the analysis of single parameter mechanism design
problems~\cite{HartlineRo2008}. This general benchmark, which bridges Bayesian analysis~\cite{Myerson1981}
from economics and worst case analysis from the theory of computer science, \ignore{is extracted through
the following template. Consider a characterization of the set of all mechanisms that are
Bayesian optimal for some i.i.d.\ distribution over possible valuations~\cite{Myerson1981}.
Define then the desired benchmark as a worst case performance benchmark that corresponds to competing
simultaneously with all on a fixed, worst case, valuation profile. It turns out
that}collides for our setting with the optimal fixed price benchmark~\cite{HartlineRo2008}.

A further justification for taking the offline fixed price auction as a
benchmark is the following. Although an online auction seems less
restricted than the offline fixed price auction, as it can assign
different players different prices, it was
shown~\cite{GoldbergHaKaSaWr2006} that the online truthful revenue is  no more
than the offline revenue.
In fact, there even exists a lower bound of $2.42$ on the competitive ratio of any
truthful online auction~\cite{GoldbergHaKaSa2004}.
Note, however, that the optimal offline revenue (or more precisely the optimal fixed price)
is unknown to the concessionaire in advance.

It is well known (see for example~\cite{Mirrlees1971}) that in order to achieve
truthfulness one can use only the set of \emph{bid independent auctions}, i.e.,
auctions in which the price offered to a bidder is independent of the bidder's own bid
value. Hence, an intuitive auction that often comes to one's mind
is the \emph{Deterministic Optimal Price (DOP) auction}~\cites{Segal2003, GoldbergHaKaSaWr2006,
AggarwalFiGoHaImSu2010}. In this auction the mechanism computes and offers each bidder
the price of an optimal offline auction for all other bids.
This auction preforms well on most bid vectors. In fact, it was even proved by
Segal~\cite{Segal2003} that if the input is chosen uniformly at random, then this auction
is asymptotically optimal. For a worst case analysis, however, it preforms very poor.
Consider, for example, an auction in which there are $n$ bidders
and only two possible bid values: $1$ and $h$, where $h \gg 1$. We denote this setting as
\emph{bi-valued} auctions. Let $n_h$ be the number of bidders
who bid $h$. Applying \textsl{DOP} on a bid vector for which
$n_h = n/h$ will result in a revenue of $n_h$ instead of
the $n$ revenue of an offline auction. This is because every ``$h$-bidder'' is offered $1$
(since $n-1 > h\cdot (n_h - 1)$) and every ``$1$-bidder'' is offered $h$ (since
$n-1 < h\cdot n_h$). Here an ``$h$-bidder'' refers to a bidder that bids $h$ and a
``$1$-bidder'' refers to a bidder that bids $1$. Therefore, the competitive ratio of \textsl{DOP}
is unbounded. Similar examples regarding the performances of \textsl{DOP}
in the bi-valued auction setting appeared already in
Goldberg~et~al.~\cite{GoldbergHaKaSaWr2006}, and in Aggarwal~et~al.~\cite{AggarwalFiGoHaImSu2010}.

Goldberg~et~al.~\cite{GoldbergHaWr2001} showed that there exist random
competitive auctions. Other works with different random competitive auctions,
competitive lower bounds, and better analysis of existing auctions were presented,
see for example~\cites{FiatGoHaKa2002, GoldbergHaKaSa2004, FeigeFlHaKl2005}.
For a survey see the work of Hartline and Karlin that appeared in~\cite{NisanRoTaVa2007}*{Chapter 13}
(Profit Maximization In Mechanism Design).
In all these works, no deterministic auction was presented. In fact,
Goldberg~et~al.~\cites{GoldbergHaWr2001, GoldbergHaKaSaWr2006} even proved
that randomization is essential assuming the auction is symmetric (aka anonymous), i.e.,
assuming the outcome of the auction does not depend on the order of the input bids.

Aggarwal~et~al.~\cites{AggarwalFiGoHaImSu2005, AggarwalFiGoHaImSu2010} later showed
how to construct from any randomized auction a deterministic, asymmetric auction
with a factor $4$ loss in the gained revenue. This result was then improved by the
authors together with Wolfovitz~\cite{Ben-ZwiNeWo2009}, but no tight derandomization
was ever presented.

\subsection{Our Results}

We introduce a tight deterministic auction for bi-valued auctions, with bid values $\{1,h\}$
and $n$ bidders. We show a polynomial time deterministic auction, which guaranty a revenue of
$\max{\{n,h\cdot n_h\}} - O(\sqrt{n\cdot h})$, where $n_h$ is just the number of
bidders that bid $h$. We then show that this bound is unconditionally optimal
by showing that every auction (including a random superpolynomial one), cannot guaranty more
than $\max{\{n,h\cdot n_h\}} - \Omega(\sqrt{n\cdot h})$. That is, there exists an
auction with no multiplicative loss and with only $O(\sqrt{n\cdot h})$ additive
loss, and every auction has at-least these losses. Let us note here,
that if we restrict ourselves to anonymous auctions (symmetric) then we
have a multiplicative loss of $\Omega(h)$ and an additive loss of
$\Omega(n/h)$ over the $\max{\{n,h\cdot n_h\}}$ revenue of the best
offline~\cites{AggarwalFiGoHaImSu2010, GoldbergHaKaSaWr2006}.

The solution for this auction is build upon a new solution for a
hat guessing game~\cite{Ben-ZwiWo2010}. The connection between digital
good auctions and hat guessing games was established by the work of
Aggarwal~et~al.~\cite{AggarwalFiGoHaImSu2005}. Here we reinforce these
connections and show a connection to a different game called the majority
game which was studied by Doerr~\cite{Doerr2005} and by Feige~\cite{Feige2004}.

\subsection{Additive Loss}
Already in the work that suggested to use competitive analysis,
namely Goldberg~et~al.~\cite{GoldbergHaWr2001}, a major obstacle arises.
It was indicated that no auction can be competitive against bids with
one high value (see Goldberg~et~al.~\cite{GoldbergHaKaSaWr2006} for details).
\ignore{Consider for example
the following example.
\begin{example}
\end{example}}
The first solution that was suggested to this problem was taking a different
benchmark as the offline auction. This different benchmark was again the maximum
single price auction, only now the number of winning bidders is bounded to
be at-least two. The term competitive was then used to indicate an auction
that has a constant ratio on every bid vector against any single price auction that
sells at-least two items.
A few random competitive auctions were indeed suggested using this definition
over the years, but, as noted before, no deterministic (asymmetric) auction was
ever found.
In fact, this was proved not to be a coincidence when
Aggarwal~et~al.~\cite{AggarwalFiGoHaImSu2005} showed that no deterministic auction
can be competitive even on this weaker benchmark.

Given this lower bound a new solution should be considered, and indeed
Aggarwal~et~al.~\cite{AggarwalFiGoHaImSu2005} suggested such. The new definition
suggested to generalize the competitive notion to include also additive losses
on top of the multiplicative ones considered before.

We argue that our results indicate that this second approach of considering also the additive loss
is more accurate, as it shows how analyzing with a finer granularity turns an
uncompetitive auction to an optimal one. We elaborate on this agenda in the Discussion
section~(\ref{conclude}).

\section{Preliminaries}
\label{prel}
A bid-vector $b\in\{1,h\}^n$ is a vector of $n$ bids.
For $b \in \{1,h\}^n$ and $i \in [n]=\{1,2,\ldots,n\}$ we
denote by $b_{-i}$ the vector which is the result of replacing the $i$th bid
in $b$ with a question mark; that is, $b_{-i}$ is the vector $(b_1, b_2,
\ldots,  b_{i-1}, ?, b_{i+1}, \ldots b_n)$.

\begin{definition}[Unlimited supply, unit demand, single item auction] An
\emph{unlimited supply, unit demand, single item auction} is a mechanism in
which there is one \emph{item} of unlimited supply to sell by an
\emph{auctioneer} to $n$~\emph{bidders}. The bidders place \emph{bids} for the
item according to their \emph{valuation} of the item.  The auctioneer then sets
prices for every bidder. If the price for a bidder is lower than or equal to
its bid, then the bidder is considered as a \emph{winner} and gets to buy the
item for its price.  A bidder with price higher than its bid does not pay nor
gets the item. The auctioneer's \emph{revenue} is the sum of the winners prices.
\end{definition}

A \emph{truthful auction} is an auction in which every bidder bids its
true valuation for the item. Truthfulness can be established through
\emph{bid-independent} auctions (see for example~\cite{Mirrlees1971}).
A \emph{bid-independent} auction is an auction for which the
auctioneer computes the price for bidder $i$ using only the vector
$b_{-i}$ (that is, without the $i$th bid).  Two models have been
proposed for describing random truthful auctions. The first, being the
\emph{truthful in expectation}, refers to auctions for which a bidder
maximizes its expected utility by bidding truthfully.  The second
model, the \emph{universally truthful} is merely a probability
distribution over deterministic auctions. Our results uses this second
definition, however, it is known that the two models collide in this setting~\cite{MehtaVa2004}.
\ignore{
\begin{definition}[Fixed price, offline auction]\label{def:offline}
The \emph{fixed price, offline auction} is the auction that on each
bid vector $b \in [h]^n$ fixes a \emph{single} price, $\alpha=\alpha(b)$ for all
bidders, so to maximize the revenue given that price. Namely, $\alpha$
is chosen such that $\sum_{b_i \geq \alpha} \alpha$ is maximized.
\end{definition}
}
\begin{definition}[General competitive auction]
Let $OPT(b)$ be the best fixed-price $($offline$)$ revenue for an $n$-bid vector $b$ with
bid values $1,\ h$. An auction $A$ is a \emph{general competitive auction} if its revenue $($expected revenue$)$
from every bid vector $b$, $P_A(b)$ is $\geq \alpha \cdot OPT(b) - o(nh)$
where $\alpha$ is a constant not depending on $n$ or $h$.
\end{definition}

\section{Bi-valued Auctions}
\label{hats}
We establish a connection between bi-valued auctions and a specific hat guessing
game known as the \emph{majority hat game}. This game was studied by
Doerr~\cite{Doerr2005} and later by Feige~\cite{Feige2004}. We use the
new results regarding this game that appeared on~\cite{Ben-ZwiWo2010}, which enable us to solve
the bi-valued auction problem optimally.

In a \emph{majority hat game} there are $n$ players, each wearing a hat colored
red or blue. Each player does not see the color of its own hat but does see the
colors of all other hats.  Simultaneously, each player has to guess the color
of its own hat, without communicating with the other players. The players are
allowed to meet beforehand, hats-off, in order to coordinate a strategy.

It was shown that the maximum number of correct guesses the players can guaranty
is no more than $\max{\{\nred,\nblue\}} - \Omega(\sqrt{n})$, where $\nred$ and
$\nblue$ are the numbers of players with red and blue hats respectively~\cites{Doerr2005, Feige2004}.
It was later proved that there exists an explicit strategy which guaranties
$\max{\{\nred,\nblue\}} - O(\sqrt{n})$ correct guesses for the players~\cite{Ben-ZwiWo2010}.

Consider \emph{bi-valued auctions}, in which there are $n$ bidders, each can select a value from $\{1,h\}$.
The auction's revenue equals the number of bidders it offers 1 plus $h$
times the number of bidders it offers $h$ if indeed their value is $h$.
\ Let $n_h(b)$ denotes the number of bidders who bids $h$ in a bid
vector $b$. Recall that the best offline revenue on vector $b$ is
$\max{\{n,h \cdot n_h(b)\}}$. In this section we will prove the following.
\begin{theorem}
\label{thm::bi}
For bi-valued auctions with $n$ bidders and values from $\{1,h\}$
\begin{enumerate}
\item	There exists a polynomial time deterministic auction that for all bid vector $b$ has revenue
\[\max{\{n,h \cdot n_h(b)\}} - O(\sqrt{n\cdot h})\]
\item There is no auction that for all bid vector $b$ has revenue
\[\max{\{n,h \cdot n_h(b)\}} - o(\sqrt{n\cdot h}).\]
\end{enumerate}
\end{theorem}

Note that the lower bound result is unconditional and applies also for
randomized superpolynomial auctions. \ We proceed with a proof for
the upper bound in the next section and a proof for the lower bound in
the adjacent one.

\subsection{An Auction}
We present next a solution to the bi-valued auction problem,
namely we show an optimal polynomial time deterministic auction.
We start again  by describing a
random auction. A derandomization will be
built later using the same methods we presented in the former section
for the hat guessing problem.\\
\ \ \\
\subsubsection{A Random Bi-valued Auction}
For a fixed input $b$, let $n_h$ be the number of $h$-bids in $b$ and
$n_h(i)$ be the number of $h$-bids in
$b_{-i}$. Let $p'(i) = \frac{h\cdot n_h(i) - n}{h\cdot \sqrt{n_h(i)}}$.
If $p'(i) \leq 0$ set $p(i) = 0$ and if $p'(i) \geq 1$ set $p(i) = 1$.
Otherwise, ($0 < p'(i) < 1$),
set $p(i) = p'(i)$.
The auction offers value $h$ for bidder $i$ with probability $p(i)$
and $1$ otherwise.

\begin{lemma}\label{lemma::bi::random}
The expected revenue of the auction described above is
$\max{\{n,h \cdot n_h\}} - O(\sqrt{n\cdot h})$
\end{lemma}
\begin{proof}
If $\exists i, p(i) \neq p'(i)$ then either $h\cdot n_h(i) \leq n$
so the auction will offer $1$ to any $1$-bidder and the revenue
will be at-least $n = \max{\{n,h \cdot n_h\}}$, or $h\cdot n_h(i) - n \geq h\cdot \sqrt{n_h(i)}$
so every $h$-bidder will be offered $h$ with probability $1 - O(1/\sqrt{n_h})$
and the expected revenue thus is at-least $hn_h\cdot(1 - 1/\sqrt{n_h}) = \max{\{n,h \cdot n_h\}} - O(\sqrt{n\cdot h})$.
Either case our auction's revenue is $\max{\{n,h \cdot n_h\}} - O(\sqrt{n\cdot h})$.
Assume now that $\forall i, p(i) = p'(i)$, note that in this case $|n - h \cdot n_h| = O(h\sqrt{n_h})$.
The expected revenue for any bid vector with $n_h$ bids of value $h$ is then:
\begin{align*}
& h\cdot n_h \frac{h\cdot (n_h - 1) - n}{h\cdot \sqrt{n_h - 1}} +
n_h \cdot (1 - \frac{h\cdot (n_h - 1) - n}{h\cdot \sqrt{n_h - 1}}) +
(n - n_h)\cdot (1 - \frac{h\cdot n_h - n}{h\cdot \sqrt{n_h}}) \\
& \ge h\cdot n_h \frac{h\cdot (n_h - 1) - n}{h\cdot \sqrt{n_h - 1}} +
n_h \cdot (1 - \frac{h\cdot (n_h - 1) - n}{h\cdot \sqrt{n_h - 1}}) +
(n - n_h)\cdot (1 - \frac{h\cdot n_h - n}{h\cdot \sqrt{n_h - 1}}) \\
&\ge h\cdot n_h \frac{h\cdot (n_h - 1) - n}{h\cdot \sqrt{n_h - 1}} +
n\cdot (1 - \frac{h\cdot n_h - n}{h\cdot \sqrt{n_h - 1}}) +
n_h \cdot (1 - \frac{h\cdot (n_h - 1) - n}{h\cdot \sqrt{n_h - 1}}) -
n_h\cdot (1 - \frac{h\cdot (n_h - 1)- n}{h\cdot \sqrt{n_h - 1}}) \\
&= h\cdot n_h \frac{h\cdot (n_h - 1) - n}{h\cdot \sqrt{n_h - 1}} +
n\cdot (1 - \frac{h\cdot n_h - n}{h\cdot \sqrt{n_h - 1}}) \\ 
&= h\cdot n_h \cdot \frac{h\cdot n_h - n}{h\cdot \sqrt{n_h - 1}} +
n\cdot (1 - \frac{h\cdot n_h - n}{h\cdot \sqrt{n_h - 1}}) -
\frac{h\cdot n_h}{\sqrt{n_h - 1}}.
\end{align*}
Observe that the sum of the first two terms in the last expression above is
$\max{\{n,h \cdot n_h\}} - O(\sqrt{n\cdot h})$. This is because
$|n - h \cdot n_h| = O(\sqrt{n\cdot h})$. The third term however,
can be absorbed also into the $O(\sqrt{n\cdot h})$, which completes the proof
of the lemma.
\end{proof}
Hence our auction's expected revenue is within an additive loss of $O(\sqrt{n\cdot h})$
from the revenue of the best offline as promised. As noted before, a derandomization for
this auction can be built using the same ideas appeared in the hat guessing game~\cite{Ben-ZwiWo2010}.
This derandomization produces an auction which has for the worst case only another additive
loss of $O(\sqrt{n\cdot h})$ over the expected revenue of the random auction.
Hence, in total, an additive loss of $O(\sqrt{n\cdot h})$ over the best
offline revenue is achieved. We sketch this derandomization here for completeness.\\
\\
\subsubsection{Derandomization}
Let $a(i) = h\cdot n_h(i) - n$
and $b(i) = h\cdot \sqrt{n_h(i)}$.
The auction computes the value offered to bidder $i$ according to the following.
\begin{enumerate}
\item Let $X(i) = \sum_j j$, where the sum ranges over all $j \ne i$ such that
the $j$th bidder bids $h$.
\item Let $Y(i) = \sum_j 1$, where the sum ranges over all $j < i$ such that
the $j$th bidder bids $h$.
\item Let $Z(i) = i + X(i) + (b(i)-1) Y(i)  \pmod {b(i)}$.
\item Offer $h$ to bidder $i$ if $Z(i) < a(i)$. Otherwise offer $1$ to the $i$'s bidder.
\end{enumerate}

Note that for the random auction whenever $p(i) = p'(i)$ (or as stated here
$a(i)/b(i) \in [0,1]$) we can define the probability that a $1$-bidder will be
offered $1$, $p_{1,1} = (1 - \frac{h\cdot n_h - n}{h\cdot \sqrt{n_h}})$,
the probability that an $h$-bidder will be
offered $1$, $p_{h,1} = (1 - \frac{h\cdot (n_h - 1) - n}{h\cdot \sqrt{n_h - 1}})$
and the probability that an $h$-bidder will be
offered $h$, $p_{h,h} = \frac{h\cdot (n_h - 1) - n}{h\cdot \sqrt{n_h - 1}}$.
\begin{lemma}
An auction that follows the above formulation gains revenue of
$n_h\cdot (h\cdot p_{h,h} + p_{h,1}) - O(\sqrt{n\cdot h})$ from all $h$-bidders.
From the $1$-bidders, the auction collects $(n-n_h)p_{1,1} - O(\sqrt{n\cdot h})$.
\end{lemma}%
\begin{proof}
Let $a(1)$ be the (identical) value $a(i)$ computed by all $1$-bidders.
In the same manner let $b(1)$, $a(h)$, $b(h)$ be the (identical) values
computed by all bidders.
The lemma follows Lemma~\ref{lemma::bi::random} and the following claim:
\begin{claim}
\
\begin{itemize}
	\item For every $b(1)$ consecutive $1$-bidders the auction will offer $h$ to $a(1)$ of them and $1$ to $b(1) - a(1)$
	\item	For every $b(h)$ consecutive $h$-bidders the auction will offer $h$ to $a(h)$ of them and $1$ to $b(h) - a(h)$
\end{itemize}
\end{claim}
\begin{proof}

Consider the $h$-bidders first. Let $1 \le i < j \le n$ be the indices
of two consecutive $h$-bidders. We have $i + X(i) = j + X(j)$ and $Y(j) - Y(i) = 1$.
Thus $Z(j) - Z(i) = b(h)-1\pmod{b(h)}$.  This implies that out of each $b(h)$ consecutive
$h$-bidders, $a(h)$ will be offered $h$ and $b(h) - a(h)$ will be offered $1$.

Next consider the $1$-bidders. Let $1 \le i < j \le n$ be the indices of
two consecutive $1$-bidders. We have $X(i) = X(j)$ and $Y(j) - Y(i) = j-i-1$.
Thus $Z(j) - Z(i) = j - i + (b(1)-1)(j-i-1) = b(1)(j-i) - b(1) + 1 \equiv 1 \pmod{b(1)}$.  This
implies that out of each $b(1)$ consecutive $1$-bidders, $b(1) - a(1)$ are offered $1$.
\end{proof}
\end{proof}
It is clear that this auction can be implemented in polynomial time as claimed, hence
the upper bound of Theorem~\ref{thm::bi} follows.\\
\begin{informalremark}
A natural critic that should arise at first glance of our ``complicated'' suggested auction
is its being ``unintuitive''. How can one explain/excuse suboptimal actions whenever $n_h\neq n/h$?
Why not deploy \textsl{DOP} in these settings?
Note, however, that the proposed auction does exactly the same. On most inputs it acts as
the \textsl{DOP} and only on inputs where $n_h \approx n/h$ it deploys the ``sophisticated''
auction. In particular, the auction sacrifices the accuracy of results whenever for the bid
vector $b$ we have that $n\leq hn_h(b) \leq n + h\sqrt{n_h(b)}$. This ``sophisticated sacrifice'',
however, results in turning an unbounded competitive auction into an optimal one.

Note also that the connection between bi-valued auctions and the majority game is in both
directions. Thus, a solution to the bi-valued auction implies a solution to the majority
game and in particular, an answer to Feige's open question~\cite{Feige2004}
\end{informalremark}

\subsubsection{A Lower Bound}
We prove optimality of the suggested auction in the previous section. For this we prove a
lower bound on the additive loss of any bi-valued auction. The lower bound is unconditional
and holds also for the expected revenue of random auctions. Furthermore, the bound does not
depend on the computation time needed for the auction.
\begin{lemma}
Let $A$ be an auction for the bi-valued $\{1,h\}$ setting and let $P_A(b)$
be $A$'s revenue on bid vector $b$. Then
$P_A(b)$ equals  $\max{\{n,h \cdot n_h\}} - \Omega(\sqrt{h\cdot n})$,
where $b$ is of size $n$ and $n_h$ is the number of bids of value $h$ in $b$.
\end{lemma}
\begin{proof}
To prove a lower bound on the difference between the maximum offline revenue,
$\max{\{n,h \cdot n_h\}}$,
and any auction we define a distribution ${\mathcal D}$ on the possible
two-values bid
vectors $\{1,h\}^n$. We then show that for any deterministic auction, the expected
revenue for a random bid vector $b$ (expectation now is with respect
to ${\mathcal D}$), is at most $P$. On the other hand, we show that the expected
revenue of the offline single price
(over the distribution ${\mathcal D}$) is at least $P + \Delta$, for some
$\Delta$. This implies (by standard averaging argument, see for example~\cite{Yao1977}),
that for any auction, (including randomized ones), there must be some vector $b$
for which the auction's revenue is $\Delta$ less than the fixed-price
offline optimal auction.

The distribution ${\mathcal D}$ in our case is quite simple: for every bidder $i
\in [n]$ independently, set $b_i=h$ with probability $1/h$ and $b_i=1$ with
probability $1- 1/h$.
Now, for every deterministic truthful auction, knowing ${\mathcal D}$, the price
for every element should better be in $[h]$, otherwise there is
another auction that assign prices in $[h]$ and achieves at least the
same revenue for every bid vector (the one that assigns $1$ for every
value less than $1$ and $h$ for every value higher than $h$). Further,
for such auction, the revenue is the sum of revenues obtained from the
$n$ bidders. Thus the expectation is the sum of expectations of the
revenue obtained from the single bidders. Since for
bidder $i$ the expectation is exactly $1$ (for any fixed $b_i$
the auction must set a constant price $\alpha \in [h]$ independent of
$b_i$. Hence for  $\alpha> 1$, the expected revenue from
bidder $i$ is $\frac{1}{h} \cdot h = 1$, and for $\alpha=1$ the
expected value is clearly $1$). We conclude that for every
deterministic truthful auction as above, the expected revenue (with
respect to ${\mathcal D}$), is exactly $n$.

We now want to prove that the expected revenue of the fixed-price
offline auction, that knows $b$, is $n+ \Omega(\sqrt{hn})$. We know,
however, the exact revenue of such auction for every bid vector
$b$. It is just $M(b)=\max{\{n,h \cdot n_h(b)\}}$, where $n_h(b)$ is
the number of $h$-bids in $b$.

Thus the expected revenue is
\begin{align}\label{eq:113}
{\mathbb E}_{{\mathcal D}}[M(b)] &:=
\sum_{i<n/h}{n \cdot \binom{n}{i} \cdot (1/h)^i \cdot (1-1/h)^{n-i}}
+ \sum_{i>n/h}{h \cdot i \cdot \binom{n}{i} \cdot
  (1/h)^i \cdot (1-1/h)^{n-i}} \\
\nonumber
& \qquad
+ \ n \cdot \binom{n}{n/h}(1/h)^{n/h}(1-1/h)^{n-n/h}.
\end{align}

To estimate this sum, it is instructive to examine the following
deterministic auction which we note before as $DOP$.
On each vector $b$, $DOP$  assigns value $h$ for every bidder $i$ for
which the number of $h$-bids in $b_{-i}$,
is at least $n/h$ (we assume $n/h$ is an integer), and $1$ otherwise.

On one side, as argued before, the expected revenue of $DOP$ with respect
to ${\mathcal D}$ is
\begin{equation}\label{eq:114}
{\mathbb E}[P_{DOP}] = n
\end{equation}

On the other hand, the same expression, is by definition,
\begin{align}\label{eq:115}
{\mathbb E}[P_{DOP}] &=
\sum_{i<n/h}{n\cdot\binom{n}{i}\cdot(1/h)^i\cdot(1-1/h)^{n-i}} +
\sum_{i>n/h}{h\cdot i \cdot \binom{n}{i} \cdot (1/h)^i \cdot (1-1/h)^{n-i}} \\
\nonumber
& \qquad + (n/h) \cdot \binom{n}{n/h}(1/h)^{n/h}(1-1/h)^{n-n/h}.
\end{align}

Comparing the expression in Equation (\ref{eq:113}) and Equation
(\ref{eq:115}), and using Equation (\ref{eq:114}), we get:
\[{\mathbb E}_D[M(b)]
= n + (n - n/h) \cdot
\binom{n}{n/h}(1/h)^{n/h}(1-1/h)^{n-n/h}.\]
Hence we conclude that the difference in expectation between offline revenue
${\mathbb E}_D[M(b)]$ and the expected revenue on any deterministic auction,
which is $n$,  is,
\[{\mathbb
  E}_D[M(b)]- n = n(1 - 1/h) \cdot \binom{n}{n/h}(1/h)^{n/h}(1-1/h)^{n-n/h}.\]
By Stirling's approximation we know that
\[\binom{n}{n/h} = \Theta\left(\frac{\sqrt{h/n}}{\sqrt{(1 - 1/h)}(1/h)^{n/h}(1-1/h)^{n-n/h}}\right).\]
Therefore, the additive loss is at least $\Omega(\sqrt{h\cdot n})$ as claimed.
\end{proof}

\section{Discussion}\label{conclude}

\emph{Bi-valued auctions} appeared in several works, such
as~\cites{GoldbergHaKaSaWr2006, AggarwalFiGoHaImSu2010, NissimSmTe2010}. We present here a connection
between these auctions and a certain \emph{hat guessing game}~\cites{Ben-ZwiWo2010, Feige2004, Doerr2005}.
The new optimal solution for this puzzle results in an optimal deterministic auction for bi-valued
auctions. Surprisingly, the establishment of the tight lower bound for
these auctions, involves with analyzing the \textsl{DOP}, the deterministic optimal
auctions for i.i.d.\ inputs.

Our recent general derandomization~\cite{Ben-ZwiNeWo2009} suffers from an additive loss of $\widetilde{O}(h\sqrt{n})$
over the expected revenue of a random auction. Aggarwal~et~al.~\cite{AggarwalFiGoHaImSu2010}
proved that every deterministic auction will suffer from an additive loss over the \emph{best
offline} auction, hence did not rule out exact derandomizations. We showed here that
\emph{every} auction (including a random one) suffers from an additive
loss of $\Omega(\sqrt{nh})$ over the best offline.
Clearly, our understanding of the additive loss is not complete yet and needs some further investigation.

Further research should ask whether there exists more cases of exact derandomization?
Is there a general exact derandomization?
Another interesting future direction, noticing that the connection between truthful auctions
and hat guessing games was not a coincidence, is to reinforce these connection, maybe with
different kind of auctions.

\bibliographystyle{plain}
\addcontentsline{toc}{chapter}{Bibliography}
\bibliography{biblio}

\begin{bibdiv}
\begin{biblist}

\bib{AggarwalFiGoHaImSu2005}{inproceedings}{
      author={Aggarwal, G.},
      author={Fiat, A.},
      author={Goldberg, A.~V.},
      author={Hartline, J.~D.},
      author={Immorlica, N.},
      author={Sudan, M.},
       title={Derandomization of auctions},
        date={2005},
   booktitle={Proceedings of the 37th annual acm symposium on theory of
  computing (stoc)},
       pages={619\ndash 625},
}

\bib{AggarwalFiGoHaImSu2010}{article}{
      author={Aggarwal, G.},
      author={Fiat, A.},
      author={Goldberg, A.~V.},
      author={Hartline, J.~D.},
      author={Immorlica, N.},
      author={Sudan, M.},
       title={Derandomization of auctions},
        date={2011},
     journal={Games and Economic Behavior},
      volume={72},
       pages={1\ndash 11},
}

\bib{Ben-ZwiNeWo2009}{inproceedings}{
      author={Ben-Zwi, O.},
      author={Newman, I.},
      author={Wolfovitz, G.},
       title={A new derandomization of auctions},
        date={2009},
   booktitle={Proceedings of algorithmic game theory, second international
  symposium (sagt)},
       pages={233\ndash 237},
}

\bib{Ben-ZwiWo2010}{inproceedings}{
      author={Ben-Zwi, O.},
      author={Wolfovitz, G.},
       title={A hat trick},
        date={2010},
   booktitle={Fun with algorithms, 5th international conference, (fun)},
       pages={37\ndash 40},
}

\bib{Doerr2005}{article}{
      author={Doerr, B.},
       title={Integral approximation},
        date={2005},
     journal={Habilitationsschrift. Christian-Albrechts-Universita"t zu Kiel},
}

\bib{Feige2004}{article}{
      author={Feige, U.},
       title={You can leave your hat on (if you guess its color)},
        date={2004},
     journal={Technical Report MCS04-03, Computer Science and Applied
  Mathematics, The Weizmann Institute of Science},
}

\bib{FeigeFlHaKl2005}{inproceedings}{
      author={Feige, U.},
      author={Flaxman, A.},
      author={Hartline, J.~D.},
      author={Kleinberg, R.~D.},
       title={On the competitive ratio of the random sampling auction},
        date={2005},
   booktitle={Internet and network economics, first international workshop,
  (wine)},
       pages={878\ndash 886},
}

\bib{FiatGoHaKa2002}{inproceedings}{
      author={Fiat, A.},
      author={Goldberg, A.~V.},
      author={Hartline, J.~D.},
      author={Karlin, A.~R.},
       title={Competitive generalized auctions},
        date={2002},
   booktitle={The 34th acm symposium on theory of computing, (stoc)},
       pages={72\ndash 81},
}

\bib{GoldbergHaKaSaWr2006}{article}{
      author={Goldberg, A.~V.},
      author={Hartline, J.~D.},
      author={Karlin, A.},
      author={Saks, M.},
      author={Wright, A.},
       title={Competitive auctions},
        date={2006},
     journal={Games and Economic Behavior},
      volume={55},
      number={2},
       pages={242\ndash 269},
}

\bib{GoldbergHaKaSa2004}{inproceedings}{
      author={Goldberg, A.~V.},
      author={Hartline, J.~D.},
      author={Karlin, A.~R.},
      author={Saks, M.~E.},
       title={A lower bound on the competitive ratio of truthful auctions},
        date={2004},
   booktitle={Stacs, 21st annual symposium on theoretical aspects of computer
  science},
       pages={644\ndash 655},
}

\bib{GoldbergHaWr2001}{inproceedings}{
      author={Goldberg, A.~V.},
      author={Hartline, J.~D.},
      author={Wright, A.},
       title={Competitive auctions and digital goods},
        date={2001},
   booktitle={Proceedings of the 12th annual acm-siam symposium on discrete
  algorithms (soda)},
       pages={735\ndash 744},
}

\bib{HartlineRo2008}{inproceedings}{
      author={Hartline, Jason~D.},
      author={Roughgarden, Tim},
       title={Optimal mechanism design and money burning},
        date={2008},
   booktitle={Proceedings of the 40th annual acm symposium on theory of
  computing, (stoc)},
       pages={75\ndash 84},
}

\bib{KoutsoupiasPi2010}{inproceedings}{
      author={Koutsoupias, E.},
      author={Pierrakos, G.},
       title={On the competitive ratio of online sampling auctions},
        date={2010},
   booktitle={Internet and network economics - 6th international workshop,
  (wine)},
       pages={327\ndash 338},
}

\bib{MehtaVa2004}{inproceedings}{
      author={Mehta, A.},
      author={Vazirani, V.~V.},
       title={Randomized truthful auctions of digital goods are randomizations
  over truthful auctions},
        date={2004},
   booktitle={Acm conference on electronic commerce},
      editor={Breese, Jack~S.},
      editor={Feigenbaum, Joan},
      editor={Seltzer, Margo~I.},
   publisher={ACM},
       pages={120\ndash 124},
}

\bib{Mirrlees1971}{article}{
      author={Mirrlees, J.~A.},
       title={An exploration in the theory of optimum income taxation},
        date={1971},
     journal={The Review of Economic Studies},
      volume={38},
      number={2},
       pages={175\ndash 208},
}

\bib{Myerson1981}{article}{
      author={Myerson, Roger~B.},
       title={Optimal auction design},
        date={1981},
     journal={Mathematics of Operations Research},
      volume={6},
      number={1},
       pages={58\ndash 73},
}

\bib{NisanRoTaVa2007}{book}{
      author={Nisan, N.},
      author={Roughgarden, T.},
      author={Tardos, {\'E}.},
      author={Vazirani, V.~V.},
       title={Algorithmic game theory},
   publisher={Cambridge University Press},
        date={2007},
}

\bib{NissimSmTe2010}{article}{
      author={Nissim, Kobbi},
      author={Smorodinsky, Rann},
      author={Tennenholtz, Moshe},
       title={Approximately optimal mechanism design via differential privacy},
        date={2010},
     journal={CoRR},
      volume={abs/1004.2888},
}

\bib{Segal2003}{article}{
      author={Segal, I.},
       title={Optimal pricing mechanisms with unknown demand},
        date={2003},
     journal={American Economic Review},
      volume={93},
}

\bib{SleatorTa1985}{article}{
      author={Sleator, D.~D.},
      author={Tarjan, R.~E.},
       title={Amortized efficiency of list update and paging rules},
        date={1985},
     journal={Commun. ACM},
      volume={28},
      number={2},
       pages={202\ndash 208},
}

\bib{Yao1977}{inproceedings}{
      author={Yao, A.},
       title={Probabilistic computations: Toward a unified measure of
  complexity (extended abstract)},
        date={1977},
   booktitle={18th annual symposium on foundations of computer science,
  (focs)},
       pages={222\ndash 227},
}

\end{biblist}
\end{bibdiv}

\end{document}